\documentclass[aps,prl,showpacs,twocolumn]{revtex4-1}

\usepackage{amsmath} 
\usepackage{graphicx}
\usepackage{amssymb} 
\usepackage{chemarr} 

\begin{document}

\title{Maximum efficiency of state-space models of molecular scale engines}

\author{Mario Einax$^{1}$}
\author{Abraham Nitzan$^{1}$}

\affiliation{
$^{1}$School of Chemistry, Tel Aviv University, Tel Aviv 69978, Israel
}

\date{\today}

\begin{abstract}
The performance of nano-scale energy conversion devices is studied in the framework of state-space models
where a device is described by a graph comprising states and transitions between them represented by nodes
and links, respectively. Particular segments of this network represent input (driving) and output processes
whose properly chosen flux ratio provides the energy conversion efficiency. Simple cyclical graphs yield
Carnot efficiency for the maximum conversion yield. We give general proof that opening a link that separate
between the two driving segments always leads to reduced efficiency. We illustrate this general result with
a simple model of an organic photovoltaic cell, where such an intersecting link corresponds to non-radiative
carriers recombination and where the reduced maximum efficiency is manifested as a smaller open-circuit voltage.
\end{abstract}


\maketitle

Modeling of molecular scale engines such as photovoltaic cells, thermoelectric devices and light emitting
diods is often done in terms of a state space description \cite{Seifert:2012}.
In such state-space models the processes underlying the device operation are described as transitions
between microscopic system states, leading to the engine description in terms of a graph comprising nodes (=states)
connected by bonds that represent transitions between them. On this network, the time evolution is determined by
the rates associated with each bonds and fluxes associated with the non-equilibrium dynamics flow along
interconnected linear and cyclical paths. Particular segments of this network represent the input
(driving) and output processes, where the corresponding rates deviate from restrictions imposed by
equilibrium thermodynamics, thus maintaining the system in a non-equilibrium state. Similar models are often used to
describe and analyze biochemical networks \cite{Hill:1966,Hill:2009,Gerritsma/Gaspard:2010}.

\begin{figure}[t!]
\centering \includegraphics[width=0.32\textwidth]{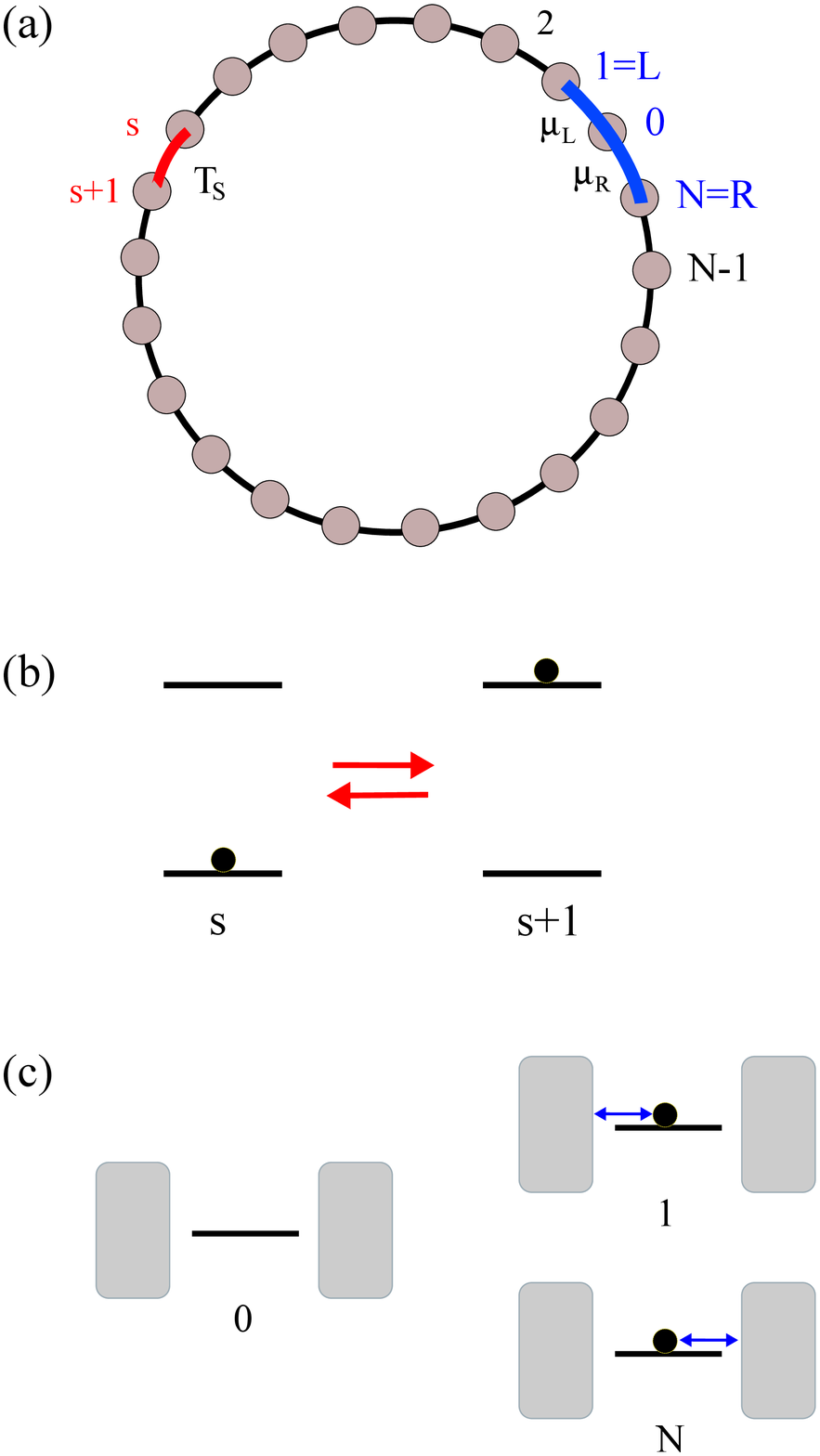}
\caption{(color online). (a) A simple graph (model SC) representing an ideal (no non-radiative losses) photovoltaic cell.
The system is driven away from equilibrium by the processes that take place at the red and blue segments.
(b) Dynamics at the red segment: States $s$ and $s+1$ are ground and excited states of a dye molecule and transition between them is driven
by the ``sun temperature'' $T_{S}$. (c) Dynamics at the blue segment. The transitions $0 \rightleftharpoons 1$
and $0 \rightleftharpoons N$ represent changes in the number of electrons on the molecule due to
its coupling to the left and right electrodes, respectively. Non-equilibrium is imposed by the
voltage difference $V$ between the two electrodes, $\mu_{R}- \mu_{L}=V$ (the electron charge $e$
is set to $1$ here and below).}
\label{fig:fig1}
\end{figure}

An example is shown in Fig.~\ref{fig:fig1} that depicts a single cycle model (henceforth referred to as model SC)
for a photovoltaic cell comprising $N$ states that represent a charge carrier occupying different sites in the system
and one state (denoted $0$) in which the charge carrier is absent. Transition rates between system states $j=0,1,\cdots,N,$
are determined by state energies $E_j$, the bias voltage $V$ and the local temperatures. The latter are taken equal to ambient
temperature $T$ everywhere except at the $s \rightleftharpoons s+1$ transition that is driven by the sun temperature $T_{S}$.
The different rates satisfy detailed balance conditions:
\begin{align}
\label{eq:DB_rates}
\frac{k_{j+1,j}}{k_{j,j+1}} &= e^{-\beta \left( E_{j+1}-E_j \right)}
\end{align}
for all $j$, $j+1$  except

\begin{align}
\label{eq:DB_rates_sun}
&\frac{k_{s+1,s}}{k_{s,s+1}} = e^{-\beta_{S} \left( E_{s+1}-E_s \right)} \qquad \qquad \qquad \textrm{(red segment)} \\
\label{eq:DB_rates_bias}
&\frac{k_{1,0}}{k_{0,1}} = e^{-\beta \left( E_{1}-\mu_{L} \right)};\, \frac{k_{N,0}}{k_{0,N}} = e^{-\beta \left( E_{N}-\mu_{R} \right)}
\;\; \textrm{(blue segment)}
\end{align}
where $k_{i,j}$ is the rate coefficient for the $j \rightarrow i$ transition
and $\beta = (k_{B} T)^{-1}$, $\beta_{\rm S} = (k_{B} T_{\rm S})^{-1}$.
The cycle affinity $\mathcal{A}$ is defined by \cite{Schnakenberg:1976}
\begin{align}
\label{eq:cycle_affinity}
\mathcal{A} &= -\ln \mathcal{K}
\end{align}
where $\mathcal{K}$ is the ratio between products of forward and backward rates
\begin{align}
\label{eq:ratio_cycle_orig}
\mathcal{K} &=\frac{k_{0,N} k_{N, N-1} \cdots k_{2,1} k_{1,0} }{k_{N,0} k_{N-1, N}\cdots k_{1,2} k_{0,1}} \, .
\end{align}

The open-circuit (OC) voltage is determined by the condition that the drivings associated with the
voltage $V$ and with the difference between $T$ and $T_{\rm S}$ balance each other so that the
steady state current through the system vanishes. In this case, the cycle affinity vanishes
\cite{Einax/Nitzan:2014}, namely $\mathcal{K}=1$. Denoting the corresponding voltage $\mu_{R}-\mu_{L}$
by $V_{\rm OC}$ and using Eqs.~(\ref{eq:DB_rates})-(\ref{eq:DB_rates_bias}) this leads to
\begin{align}
\label{eq:OC_voltage}
\frac{V_{\rm OC}}{\Delta E} = 1-\frac{T}{T_{S}}\equiv \eta_{C}\,; \quad \Delta E= E_{s+1}-E_{s} \, .
\end{align}
The l.h.s is $\lim_{J\rightarrow 0} \eta (J)$ where  $\eta(J) = V(J)/\Delta E$ is the thermodynamic
(``internal'') efficiency of the device and $J$ is the steady state current (same through all links).
Equation (\ref{eq:OC_voltage}) thus yields the Carnot expression $\eta_{C}$ for the efficiency in the zero power limit.
It also highlights the OC voltage as an important attribute in the quest to understand cell performance \cite{Vandewal/etal:2009,Kirchartz/etal:2009a,Potscavage/etal:2009,Nayak/etal:2011}.
In Ref.~\onlinecite{Einax/etal:2011} we have derived two important generalizations of this result: First,
if the transition $s \rightleftharpoons s+1$ results from the combinations of radiative and non-radiative processes
whose forward/backward ratios are determined by temperatures $T_{\rm S}$ and $T$, respectively,
Eq.~(\ref{eq:OC_voltage}) is replaced by a similar expression in which $T_{\rm S}$
is substituted by an effective temperature\cite{note:PRL1}
\begin{align*}
T_{\rm eff}&=T_{\rm S} \left( 1+\frac{k_{\rm B}T}{\Delta E} \ln \left[ \frac{1+\rho}{1+\rho \exp[-(\beta-\beta_{\rm S})\Delta E]}\right] \right)^{-1}\\
&\simeq T_{\rm S}\left( 1+\frac{k_{\rm B}T}{\Delta E}\ln[1+\rho]\right)^{-1}\, ,
\end{align*}
where $\rho=k_{s,s+1}^{\rm NR}/k_{s,s+1}^{\rm R}$
is the ratio between the non-radiative and radiative relaxation rates.
Second, if one of the cycle processes involves charge separation that comes at an energy cost (exciton binding energy) $E_{B}$,
Eq.~(\ref{eq:OC_voltage}) is replaced by \cite{Einax/Nitzan:2014}
\begin{align}
\label{eq:Voc_BHJ-OPV}
\frac{V_{\rm OC}}{\Delta E} &= \eta_{C} - \frac{E_{B}}{\Delta E} \,.
\end{align}
The later situation characterizes the operation of organic photovoltaic cells \cite{Giebink/etal:2011,Koster/etal:2012,Gruber/etal:2012}.

Model SC and Eqs.~(\ref{eq:cycle_affinity})-(\ref{eq:Voc_BHJ-OPV}) constitute a convenient framework for
analyzing the OC voltage of photovoltaic cells, and, as shown in Ref.~\onlinecite{Einax/Nitzan:2014},
can be used as a starting point for studying optimal performance in finite power operations, provided that
information on the system electronic structure and dynamics can be incorporated into such state-space model.
However, this single cycle model is oversimplified; realistic state-space models contain intersecting pathways
that makes the analysis that leads to Eqs.~(\ref{eq:cycle_affinity})-(\ref{eq:OC_voltage}) more involved.
Indeed, in state-space models of photovoltaic cells such intersecting pathways represent non-radiative losses,
e.g., electron-hole recombination processes that are expected to reduce efficiency.

\begin{figure}[t!]
\centering \includegraphics[width=0.35\textwidth]{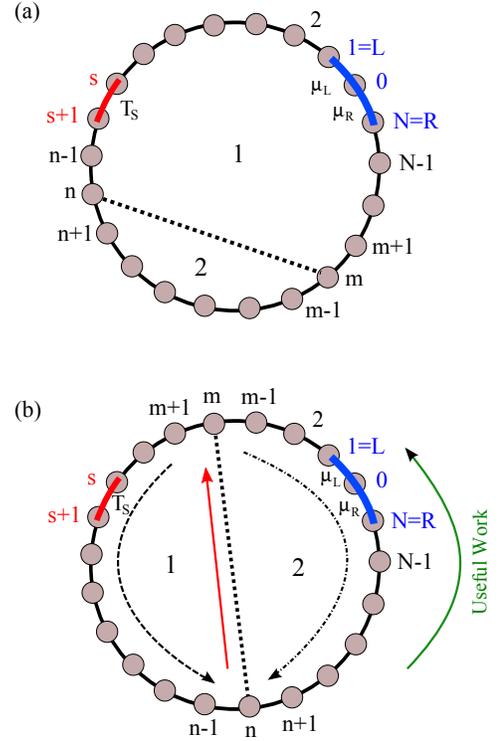}
\caption{(color online). Models characterized by intersected graphs - an new pathway is added between
sites $m$ and $n$. (a) The intersecting pathway does not separate the two driving
(red and blue) segments. (b) The bond $n-m$ separates between the two driving segments which now
sit on different subcycles.}
\label{fig:fig2}
\end{figure}

As graphical structures, such intersecting pathways divide the cycle of model SC into separate cycles
(see Fig.~\ref{fig:fig2}), making it possible to have steady states in which the current in the
power-extracting segment vanishes while internal currents exist elsewhere in the system.
While intuitively we expect lower efficiencies in such situations, no general proof for this is available.

In this Letter we consider two generalizations of model SC shown in Fig.~\ref{fig:fig2}.
In one [Fig.~\ref{fig:fig2}(a)], an intersecting pathway divides the system cycle into
two subcycles without separating between the two driving segments.
In the second [Fig.~\ref{fig:fig2}(b)], the driving segments are positioned on different cycles.
We show that 
\emph{(a)} Intersecting pathways of the type shown in Fig.~\ref{fig:fig2}(a) do not affect the OC voltage,
hence their zero power operation is characterized by the Carnot efficiency.
\emph{(b)} Intersecting pathways of the type shown in Fig.~\ref{fig:fig2}(b) lead to zero
power efficiency smaller that the Carnot value, manifested for photovoltaic cells be a lower OC voltage.

We start with Model SC and take note of the two sites, $m$ and $n$, between which a new pathway will be enabled.
The rate ratio, Eq.~(\ref{eq:ratio_cycle_orig}), can obviously be written as $\mathcal{K}=\bar{\mathcal{K}}_1 \bar{\mathcal{K}}_2$
where $\bar{\mathcal{K}}_1=k_{n,n-1}\cdots k_{m+1,m}/k_{n-1,n}\cdots k_{m,m+1}$ and
$\bar{\mathcal{K}}_2=k_{m,m-1}\cdots k_{n+1,n}/k_{m-1,m}\cdots k_{n,n+1}$ are ratios of rate products that correspond to
the arcs $1$ and $2$ in Fig.~\ref{fig:fig2}.
Now enable the pathway between sites $n$ and $m$, that is, make $k_{n,m}$ and $k_{m,n}$ different from zero.
The graph is now divided into two subcycles, $1$ and $2$ in addition to the original cycle. The corresponding cycle
rate-product ratios are
\begin{align}
\label{eq:ratios_sub-cycles}
\mathcal{K}_1 &= \bar{\mathcal{K}}_1 \frac{k_{m,n}}{k_{n,m}}\,; \quad \mathcal{K}_2 = \bar{\mathcal{K}}_2 \frac{k_{n,m}}{k_{m,n}}
\end{align}
so that $\mathcal{K}_1 \mathcal{K}_2 = \bar{\mathcal{K}}_1 \bar{\mathcal{K}}_2$.

Consider next the situation shown in Fig.~\ref{fig:fig2}(a), where the two driving segments are on the same subcycle $1$.
Starting from the model SC in the OC situation so that $\mathcal{K}=\mathcal{K}_1\mathcal{K}_2=1$, enabling the $n-m$ link
does not lead to current on this bond. To show this note that since all rates on arc $2$ satisfy Eq.~(\ref{eq:DB_rates}),
we have $\bar{\mathcal{K}}_2=e^{-\beta(E_m-E_n)}=k_{m,n}/k_{n,m}$. Equation~(\ref{eq:ratios_sub-cycles}) then yields
$\mathcal{K}_2=1$ and therefore $\mathcal{K}_1=1$. Consequently, connecting the sites $n$ and $m$
will not create any extra current and subsequently does not affect the open-circuit voltage.

In the situation of Fig.~\ref{fig:fig2}(b), each of the subcycles created by enabling the $n-m$ pathway contains a non-equilibrium element:
driving by the sun in subcycle $1$ and the driving by the voltage in subcycle $2$. Focusing on the latter and using Eqs.~(\ref{eq:DB_rates}) and (\ref{eq:DB_rates_bias}) yields
\begin{align}
\label{eq:barA2_caseB}
\bar{\mathcal{K}}_2 &= e^{-\beta (E_{m}-E_{n})} e^{-\beta \Delta\mu}= \frac{k_{m,n}}{k_{n,m}} e^{-\beta \Delta\mu} \, ,
\end{align}
where $\Delta \mu=\mu_{R}-\mu_{L}$. Equation~(\ref{eq:ratios_sub-cycles}) then yields $\mathcal{K}_2 = \exp[-\beta \Delta\mu]$.
This indicates that with the $m-n$ link enabled the system cannot establish a state where currents through all links in subcycle $2$
vanish unless $\Delta\mu=0$.

To understand the consequence for the open-circuit voltage, let us set $k_{n,m}=\lambda \bar{k}_{n,m}$ and $k_{m,n}=\lambda \bar{k}_{m,n}$
and evaluate the derivative of the OC voltage, $V_{\rm OC}$, with respect to $\lambda$ at $\lambda=0$. $V_{\rm OC}$ is the value of
$\Delta\mu=\mu_{R}-\mu_{L}$ for which the current between the electrodes, (the blue segment in Figs.~\ref{fig:fig1} and \ref{fig:fig2}),
therefore across all links in arc $1$, vanishes; namely $k_{j,j+1}P_{j+1} = k_{j+1,j} P_j$ for all this links.
From Eqs. (\ref{eq:DB_rates}) and (\ref{eq:DB_rates_bias}) it follows that under OC conditions,
the probabilities to find the system in states $n$ and $m$ are related to $V_{\rm OC}$ by
\begin{align}
\label{eq:ratio_PmPn}
\frac{{P}_m}{{P}_n} &= e^{-\beta (V_{\rm OC} + E_{m}-E_{n})}  \, .
\end{align}
It follows that
\begin{align}
\label{eq:derivation_lambda}
\left( \frac{d V_{\rm OC}}{d\lambda}\right)_{\lambda=0} = -c \left(\frac{d }{d\lambda} \frac{\bar{P}_m}{\bar{P}_n} \right)_{\lambda=0}
\end{align}
where $c$ is a positive number and $P_j=\bar{P}_j e^{-\beta E_j}$.
The derivative on the r.h.s is negative if ($\bar{P}_m>\bar{P}_n)_{\lambda=0}$, i.e. $V_{\rm OC}<0$,
and negative in the opposite case, ($\bar{P}_m<\bar{P}_n)_{\lambda=0}$; $V_{\rm OC}>0$
(this can be inferred from the current expression,
$J_{n\leftarrow m} = \lambda \left( \bar{k}_{n,m} P_m - \bar{k}_{m,n} P_n \right)= \lambda \bar{k}_{n,m} e^{-\beta E_m} \left(\bar{P}_m - \bar{P}_n \right)$.
Hence
\begin{align}
\label{eq:negative_derivative_lambda}
\left( \frac{d |V_{\rm OC}|}{d\lambda}\right)_{\lambda=0} < 0\, .
\end{align}
Enabling the $m-n$ link in the scheme of Fig.~\ref{fig:fig2}(b) thus results in reduction of the OC voltage,
namelt of the maximum thermodynamic efficiency.

\begin{figure}[t!]
\centering \includegraphics[width=0.4\textwidth]{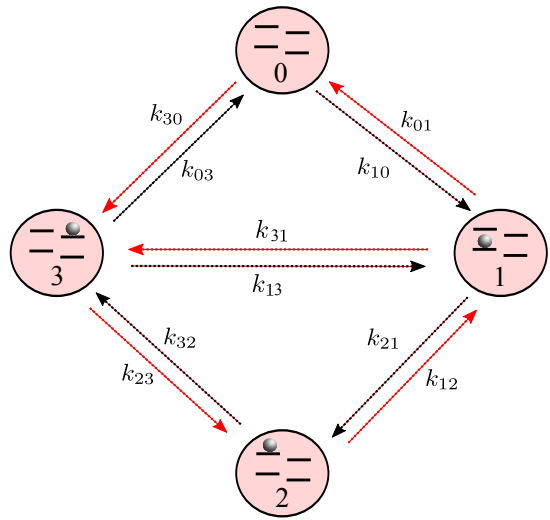}
\caption{(color online). A $4$-level spinless model of an organic photovoltaic cell\cite{Einax/etal:2011,Einax/etal:2013}.
The states are: $0$ - no excess electron, $1$ - excess electron in the higher occupied molecular orbital (HOMO) of the donor,
$2$ - Following an optical transition, electron is on the otherwise lower unoccupied molecular orbital (LUMO) of the donor (an exciton state).
$3$: Following charge separation, the electron has moved to the acceptor LUMO.
The transitions $3 \rightleftharpoons 0$ and $0 \rightleftharpoons 1$ include electron transfer to the right and left electrode,
respectively, $1 \rightleftharpoons 2$ is the optical excitation process and $2 \rightleftharpoons 3$ is the exciton dissociation
process (see Ref.~\onlinecite{Einax/etal:2011} for further details). The process
$3 \rightleftharpoons 1$ that divides the cycle to two subcycles represents radiationless recombination.}
\label{fig:fig3}
\end{figure}

To illustrate this general observation for a particular system, we consider a simple $4$-level model
for an organic photovoltaic cell comprising a donor acceptor (D-A) complex
placed between two electrodes (see Fig.~\ref{fig:fig3} and Supplementary Material~\onlinecite{note:supp_mat}).
The $D$ and $A$ species are characterized by their HOMO levels, $\varepsilon_{\rm D_H}$, $\varepsilon_{\rm A_H}$,
and LUMO levels, $\varepsilon_{\rm D_L}$, $\varepsilon_{\rm A_L}$ that translate into three relevant energy differences
$\Delta E = \varepsilon_{\rm D_L}- \varepsilon_{\rm D_H}$ (the optical gap),
$\Delta \epsilon = \varepsilon_{\rm D_L}- \varepsilon_{\rm A_L}$ (interfacial LUMO-LUMO gap), and
$E_{DA,g}=\varepsilon_{\rm A_L}-\varepsilon_{\rm D_H}$ (effective band gap).
Charge carriers can enter and leave the system only by transitions between the donor HOMO level and the electrode $L$,
and between the acceptor LUMO level and the electrode $R$. These electrodes are represented by free-electron reservoirs
at chemical potentials $\mu_K$ ($K=L;R$). An important feature of this model is that a thermal fluctuation is needed
to overcome the exciton binding energy $E_{B}$ of an interfacial charge transfer exciton to generate a free charge carrier.
With some simplifying assumptions, the system can be described by a kinetic scheme involving four states.
A detailed description of this model and the underlying kinetic scheme was given in Ref.~\onlinecite{Einax/etal:2011}.

An additional link between states $1$ and $3$ correspond to non-radiative recombination that returns a free electron on the
acceptor back to the donor ground state. In the absence of this process,
the open-circuit voltage $V_{\rm OC} (\lambda=0)$ is found to be given by Eq.~(\ref{eq:Voc_BHJ-OPV}).
The maximum thermodynamic efficiency is correspondingly $\eta_{C}-E_{B}/\Delta E$,
highlighting the loss associated with the exciton binding energy.

Repeating the cycle analysis when the rates $k_{13}$, $k_{31}$ are finite \cite{note:supp_mat}
yields the following expression for the open-circuit voltage
\begin{align}
\label{eq:Voc_loss}
V_{\rm OC} (\lambda) &= \Delta E \eta_{C} - E_{B} -
k_{B} T \ln \frac{\displaystyle 1+k_{13} \frac{k_{12}+k_{32}}{k_{12}k_{23}} }{\displaystyle 1+k_{31} \frac{k_{12}+k_{32}}{k_{21}k_{32}} }\, ,
\end{align}
setting  $k_{13}=\lambda \overline{k}_{13}$ and $k_{31}=\lambda \overline{k}_{31}$.
and taking the derivative with respect to $\lambda$ we find, using also Eqs.(S5-S8)
\begin{align}
\label{eq:derivation_lambda_BHJ-OPV}
\left( \frac{d V_{\rm OC} }{d \lambda}\right)_{\lambda=0} &=- k_{B} T \frac{\overline{k}_{13} (k_{32} + k_{12})}{k_{12} k_{23}}
\left( 1 - e^{-\beta (V_{\rm OC})_{\lambda=0}}\right) \, .
\end{align}
where $(V_{\rm OC})_{\lambda=0}$ is given by (\ref{eq:Voc_BHJ-OPV}).
This is an explicit statement of the inequality (\ref{eq:negative_derivative_lambda}).

In conclusion, we have presented a state-space network representation scheme of molecular scale engines
and applied it to models of photovoltaic cells. A general argument based on cycle analysis yields
conditions under which the opening of intersecting pathways in an otherwise cyclical graph  leads
to a decrease in the maximum efficiency. In the application to the operation of a photovoltaic cell
such intersecting link often corresponds to a carrier recombination process, and its consequence
is manifested in a reduced open-circuit voltage. It should be emphasized that similar analysis can
be carried out for more complex models that include, for example, polaron formation and hot exciton dissociation,
as well as for other types of nano-scale engines, provided that they can be modeled by kinetic transitions in the system state-space.
As described elsewhere \cite{Einax/Nitzan:2014}, such an approach can be used for analysis of
optimal performance under finite power operation, although with the unavoidable loss
of the generality associated with a thermodynamic description.

\end{document}